\documentclass[aps,prb,twocolumn,shownopacs,superscriptaddress]{revtex4}
\usepackage{amsmath,graphicx,amsbsy,amssymb,amsmath,epsfig,latexsym,wasysym,pifont,mathrsfs,natbib}
\usepackage{float}
\newfloat{widefig}{thp}{lop}
\usepackage{graphicx}
\usepackage{comment}
\usepackage{verbatim}
\usepackage{multirow,array}
\usepackage{hyperref}
\usepackage{subfigure}
\usepackage{color}
\begin{document}

\title{Half-metallicity in quaternary Heusler alloys with 3$d$ and 4$d$ elements: observations and insights from DFT calculations. }

%%%%%%%%%%%%% \iffalse

\author{Srikrishna Ghosh}
%\email{srikrishna@iitg.ernet.in}
\affiliation{Department of Physics, Indian Institute of Technology Guwahati, Guwahati-781039, Assam, India.} 
%\author{Ashis Kundu}
%\affiliation{Department of Physics, Indian Institute of Technology Guwahati, Guwahati-781039, Assam, India.}
\author{Subhradip Ghosh}
\email{subhra@iitg.ac.in}
\affiliation{Department of Physics, Indian Institute of Technology Guwahati, Guwahati-781039, Assam, India.}

\date{\today}
%=========================================================================
%========================= Abstract ==================================
\begin{abstract}  
 In this work, we provide important insights into the evolution of half-metallicity in quaternary Heusler alloys. Employing {\it ab initio} electronic structure methods we study 18 quaternary Heusler compounds having the chemical formula CoX$^\prime$Y$^\prime$Al, where  Y$^\prime$ = Mn, Fe;  and X$^\prime$ a 4$d$ element. Along with the search for new materials for spintronics applications, the trends in structural, electronic, magnetic properties and Curie temperature were investigated. We have made comparative studies with the compounds in the quaternary series CoX$^{\prime}$Y$^{\prime}$Si with X$^{\prime}$ materials from 3$d$ and 4$d$ transition metal series in the periodic table. We observe that the half-metallic behaviour depends primarily on the crystal structure type based on atomic arrangements and the number of valence electrons. As long as these two are identical, the electronic structures and the magnetic exchange interactions bear close resemblances. Consequently, the materials exhibit identical electronic properties, by and large. We analysed the roles of different transition metal atoms in affecting hybridisations and correlated them with the above observations. This work, therefore, provides important perspectives regarding the underlying physics of half-metallic behaviour in quaternary Heusler compounds which goes beyond specifics of a given material. This, thus, paves way for smart prediction of new half-metals. This work also figures out an open problem of understanding how different ternary Heuslers with different electronic behaviour may lead to half-metallic behaviour in quaternary Heuslers with 4$d$ transition metal elements.
\end{abstract}

\maketitle
%=========================================================================
%========================= Introduction ==================================
\section{INTRODUCTION}

Electronic devices with metals, semiconductors and insulators as the constituting components, have been integral part of the transient transformation of society for over a century. Information processing in conventional electronic devices is based only on the charge of the electrons. Spin electronics or spintronics, uses the spin of electrons, as well as their charge, to process information.  Significant advantages are achieved if the target properties are observed in classes of compounds with simpler crystal structures as that  can provide insights into the structure-property relationships in a tractable way. After the exciting discovery of half metallic ferromagnetisim \cite{Groot1983} in Heusler systems,  Heusler intermetallics  have been exhaustively studied for its potential applications not only in magneto-electronics but also for  possibilities to discover other novel magnetic phenomena  in this group of compounds. Recent discovery of Spin Gapless Semiconductors (SGS)\cite{wang2008sgs,OuardiPRL13} in structures of Heusler variety further boosted it's prospect as a sought after material class for spintronics applications. 

Full Heuslers alloys (L2$_1$), X$_{2}$X$^{\prime}$Z  are largest in number among the explored intermetallics, where X and X$^{\prime}$ are transition metal atoms and Z is  a main group element sitting within four interpenetrating fcc sub-lattices. The  occupancies of the different  Wyckoff positions by the constituent atoms led to two different kind of Heusler structures namely the ``regular'' and the ``inverse'' Heusler structures with space group $Fm\bar{3}m$ and $F\bar{4}3m$ respectively (space group number 225 and  216 respectively). It is obvious that with four different atoms occupying all the Wyckoff positions ( symmetric site pair $4a(0,0,0)$, $4b(1/2, 1/2, 1/2)$ and  $4c(1/4, 1/4, 1/4)$, $4d(3/4, 3/4, 3/4)$ ) a quaternary Heusler structure known as LiMgPdSn type or Y-type , is obtained. A simple permutation of the site occupancies by different atoms reveals that there are three possible Y-type structures. The three possible crystal structure of equiatomic quaternary Heusler alloys with chemical formula XX$^\prime$Y$^\prime$Z are shown in Fig \ref{Crystal1}. Due to greater flexibilities arising out of the choices for magnetic components and their arrangements on different lattice sites the quaternary Heusler alloys are found to be indispensable  for the exploration of potential new functional materials. However, compared to the ternary compounds, the exploration of quaternary Heuslers are recent and relatively fewer in number. Nevertheless, there already are a handful of quaternary compounds exhibiting properties useful for spintronics devices \cite{Alijani1PRB11,AlijaniPRB11,BainslaJAC15,BainslaPRB215,XuEPL13,DaiJAP09,CoFeMnSi_aftab}.

The ternary and quaternary Heusler alloys that have been extensively investigated mostly have 3$d$  transition metals as magnetic components \cite{AlijaniJAP13, MeshcheriakovaPRL14,VajihehJPCM12,GalanakisJAP14, XieCMS15, EndoJAC12, GrafPSSC11, EnamullahPRB15, WollmannPRB14}. It is well known that the interesting properties of Heusler alloys like the half-metallic behaviour arise mainly due to the  the hybridisations of the unfilled 3$d$ shells of different magnetic components present in a structure. It is, therefore, expected that Heusler compounds with 4$d$ transition elements can expand the scope of discovering new materials for spintronics applications. Currently, there are very few investigations on Heusler compounds having 4$d$ elements with regard to half-metallic systems, even for ternary ones. For quaternary systems, they are even less. Moreover, the investigations into the quaternary half-metals are not systematic and do not attempt in unraveling the details of electronic structures so as to understand the  trends across different series.  Recently few new quaternary half-metallic ferromagnets with large Curie temperatures were reported where one out of the three magnetic components was a 4$d$ element. This work on CoX$^{\prime}$Y$^{\prime}$Si series (X$^{\prime}$ covers all 9 elements in the row of 4$d$ transition metals and Y$^{\prime}$=Fe,Mn) \cite{NSRP} made an attempt to go beyond only reporting new half-metals, by performing an in-depth study on the relative importances of 3$d$ and 4$d$ elements affecting half-metallic behaviour.    

In a bid to systematically understand the physics of half-metallic behaviour in quaternary Heusler compounds where one of the magnetic elements is from the 4$d$ series of periodic table, in this work, we have computed the structural, electronic and magnetic properties of compounds in two quaternary Heusler series, CoX$^{\prime}$MnAl and CoX$^{\prime}$FeAl, where X$^{\prime}$ is an element with 4$d$ electrons. Apart from this, we perform a comparative study of the four series CoX$^{\prime}$Y$^{\prime}$Z where Z=Al,Si; Y$^{\prime}$=Fe,Mn and X$^{\prime}$ are the 4$d$ elements. Apart from the quest for new half-metals, following are the issues we attempt to address: 
\begin{figure}[H]
\centering

\subfigure[]{\includegraphics[scale=0.025]{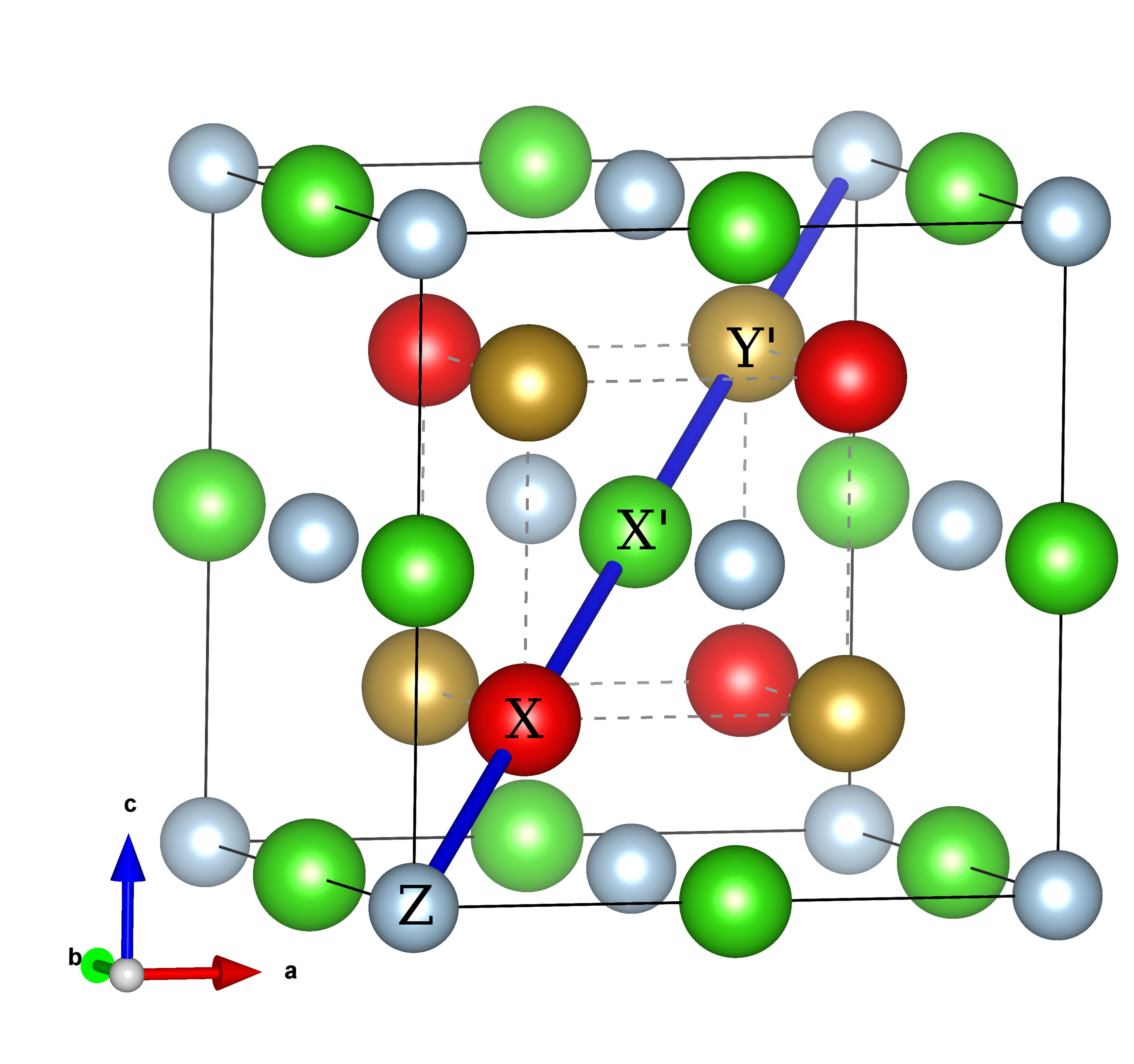}}
\subfigure[]{\includegraphics[scale=0.025]{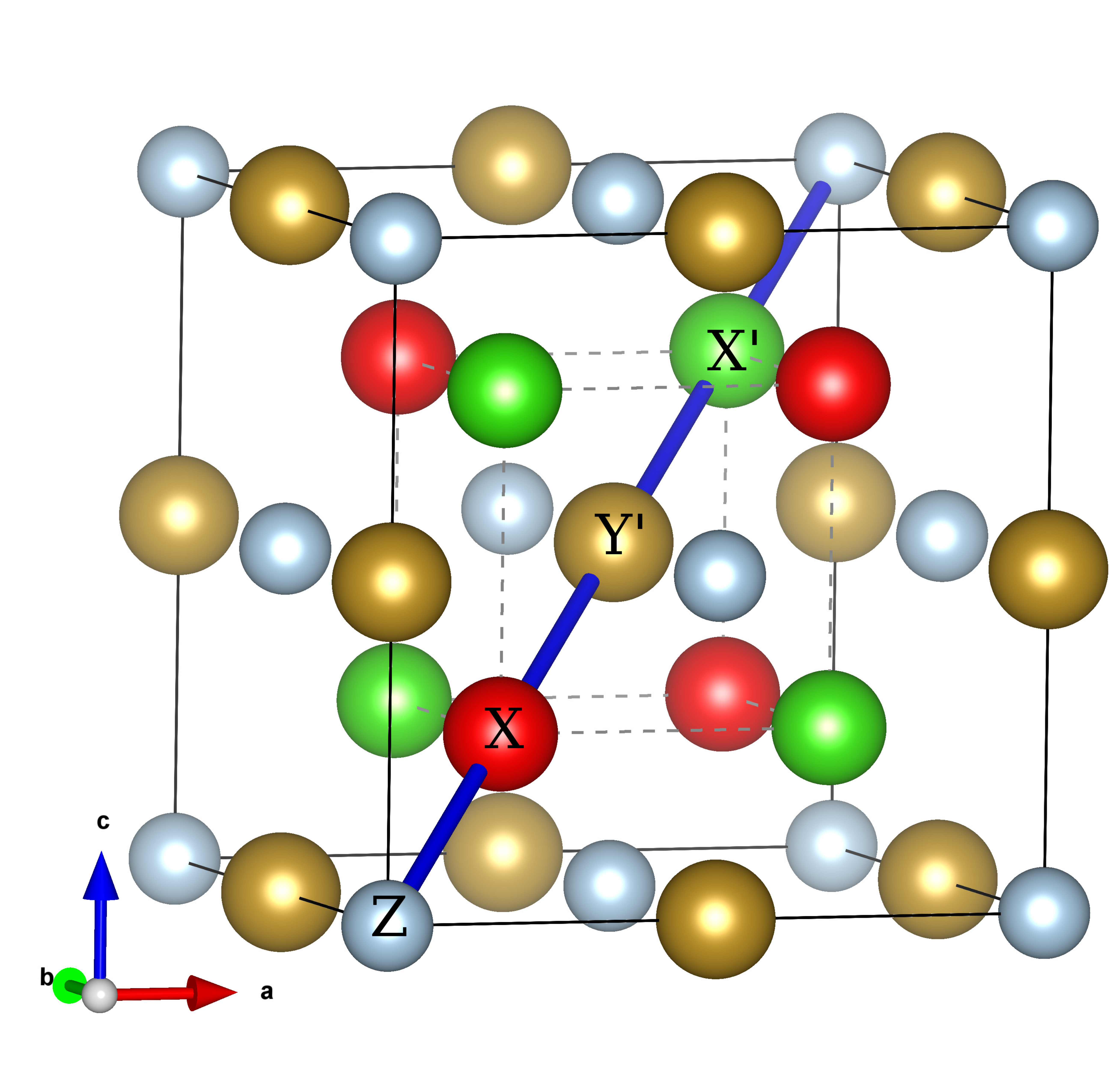}}
\subfigure[]{\includegraphics[scale=0.025]{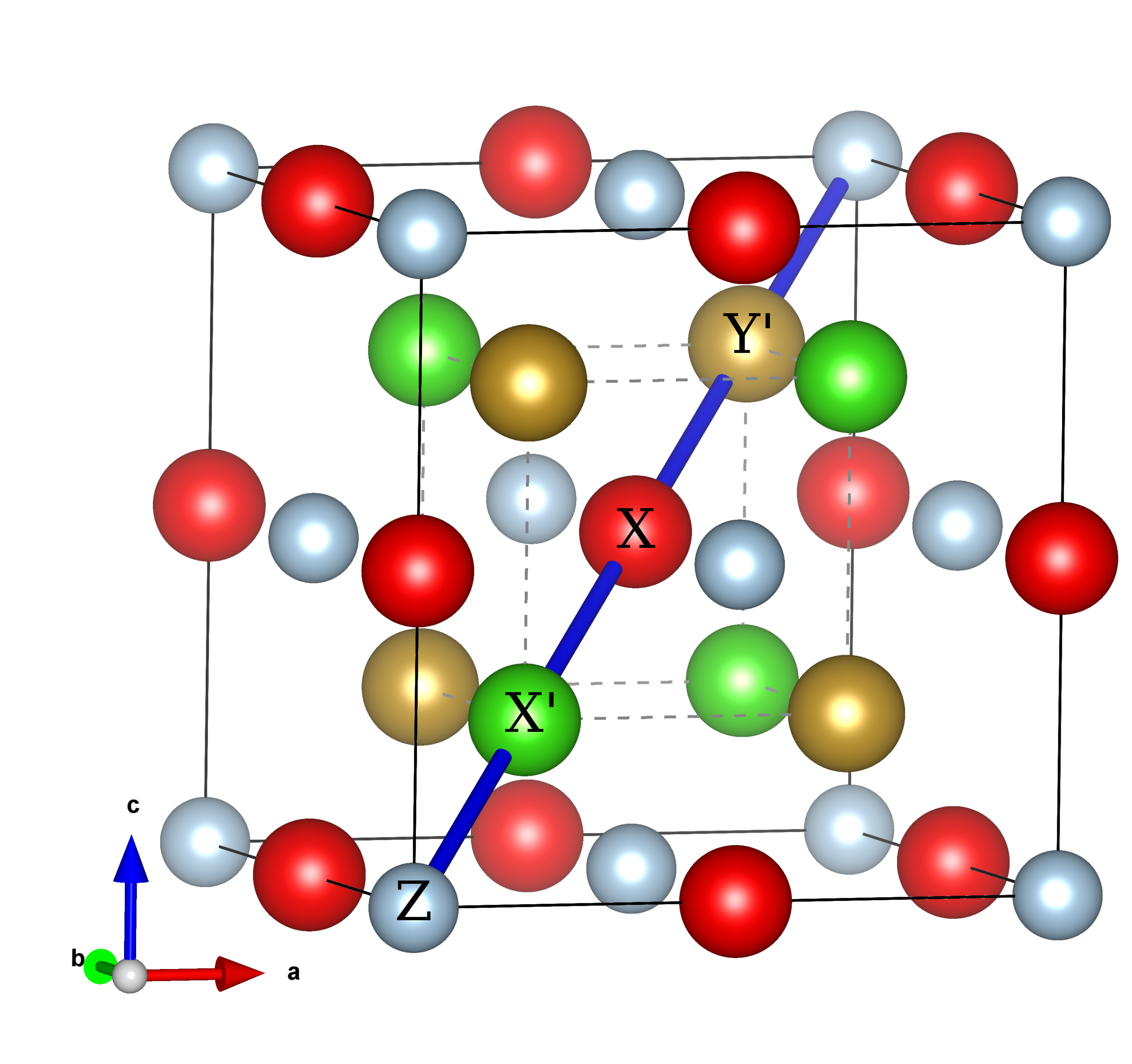}}

\caption{Crystal structures of quaternary Heusler compounds XX$^{\prime}$Y$^{\prime}$Z: ({\bf a}) structure type T$_{\text I}$ (X and Y$^{\prime}$ are in symmetric positions 4c and 4d respectively; 4a and 4b sites are occupied by Z and X$^{\prime}$) ({\bf b}) structure type T$_{\text{II}}$ (X and X$^{\prime}$ are in symmetric positions 4c and 4d; 4a and 4b sites are occupied by Z and Y$^{\prime}$.) ({\bf c}) structure type T$_{\text{III}}$(X$^{\prime}$ and Y$^{\prime}$ are in symmetric positions 4c and 4d). }
\label{Crystal1}
\end{figure}

%%%%%%%%%%%%%%%%%%%%%%%%%%%%%%%%%%%%%%%%%%%%%%%%%%%%%%%%%%%%%%%%%%%%%%%%%%%%%%%%%%%%%

(1) to understand how the similarities and differences between the arrangements of atoms in various crystallographic sites in the ground states, and the electronic structures, influence the possibility of half-metallic behaviour in quaternary compounds with 4$d$ elements.

(2) to understand the key factors influencing half-metallic behaviour in quaternary compounds by qualitative comparison between compounds where all magnetic atoms are from 3$d$ atoms and where one of the magnetic atoms are from 4$d$ series. 

(3) to observe whether half-metallic behaviour is transferrable to daughter quaternary compounds with 4$d$ elements whose parent compounds are ternary Heuslers. This is important in the light of a recent work on Co$_{2}$, Fe$_{2}$,Mn$_{2}$ based ternary systems where the second magnetic element is one of the 4$d$ transition metals, and the half-metallic behaviour was observed in few of them which could be understood in terms of the atomic arrangements, the number of valence electrons and the electronic structures as a consequence of these \cite{srikrishna1}.

(4) to find out the role of $4d$ transition metal elements by successively changing it along the series of $4d$ transition metal elements so that the total number of valance electron changes continuously, on the properties of compounds where one of the $3d$ elements remains fixed with the other changes.

The paper is organised as follows. In the next section we present computational details and the methods,. The discussions of the calculated results in details followed by a conclusion and future outlook is given in the subsequent  subsection. %Computational details and the methods, used in this work are given at the end.
 
%%%%%%%%%%%%%%%%%%%%%%%%%% End of Introduction %%%%%%%%%%%%%%%%%
%=======================================================================================

%========================= Computational Details ==================================

\section{Computational Details}

 Using Vienna Ab-initio Simulation Package (VASP) \cite{PAW94,VASP196,VASP299}, we have used spin-polarised DFT based projector augmented wave methods for calculating electronic structure with Generalized Gradient Approximation (GGA) \cite{PBEGGA96}. 450 eV energy cut-off and a Monkhorst-Pack \cite{MP89} 25 x 25 x 25 $k$-mesh were used for self-consistent calculations while we used a larger 31 x 31 x 31 $k$-mesh for the densities of sates. The energy and the force convergence criteria were set to $10^{-6}$ eV and $10^{-2}$ eV/\AA\  respectively. We checked our results by turning on the spin-orbit coupling as systems had 4$d$ electrons. However, the results did not change.
  
 The magnetic pair exchange parameters were calculated with multiple scattering Green function formalism as implemented in SPRKKR code\cite{EbertRPP11}. {%The spin part of the Hamiltonian is mapped to a Heisenberg model%}
 The Heisenberg Hamiltonian is used for the spin interaction of sites using the following formula
\begin{eqnarray}
H = -\sum_{\mu,\nu}\sum_{i,j}
J^{ij}_{\mu\nu}
\mathbf{e}^{i}_{\mu}
.\mathbf{e}^{j}_{\nu}
\end{eqnarray}
where $\mu$, $\nu$ refers to different sub-lattices; \emph{i}, \emph{j} indices are used to denote the atomic positions and $\mathbf{e}^{i}_{\mu}$ is the directional unit vector along the direction of magnetic moments at site \emph{i} belonging to sub-lattice $\mu$. Using the formulation  of Liechtenstein {\it et al.}\cite{LiechtensteinJMMM87},  $J^{ij}_{\mu\nu}$ are calculated from the energy differences due to infinitesimally small orientations of a pair of spins. We have used full potential spin polarised scaler relativistic Hamiltonian with angular momentum cut-off $l_{max} = 3$ to calculate the energy differences by the SPRKKR code. A uniform $k$-mesh of 22 x 22 x 22 is used  for Brillouin zone integration. The Green's functions were calculated for 32 complex energy points distributed on a semi-circular contour. For the self-consistent cycles, the energy convergence criterion was set to 10$^{-6}$ Ry . The  calculated $J^{ij}_{\mu\nu}$ were used further to compute the Curie temperatures T$_{c}$ using the mean field approximation\cite{SokolovskiyPRB12}.  

%\clearpage
%\newpage
%%%%%%%%%%%%%%%%%%%%%%%%%% End of Computational Details %%%%%%%%%%%%%%%%%
%=======================================================================================
%========================= Results and Discussions ==================================
\section{Results and Discussions}

\subsection{Structural Properties}
As shown in Fig. \ref{Crystal1} there are three possible crystal structures for a quaternary Heusler alloy XX$^\prime$Y$^\prime$Z.  We first optimised the ground state structures for each of the compounds by computing their total energies in each of the three structures. In Table \ref{table1} we have tabulated the calculated lattice constants, the formation energies and total magnetic moments for each of the 18 quaternary Heusler alloys in their respective ground state. We found that  structure type T$_{\text{III}}$ is energetically always higher. 
The reason behind none of the compounds crystallising in T$_{\text{III}}$ is explained in Ref. \onlinecite{NSRP}. This, along with the structure types in which each of the compounds investigated crystallise, can be explained on the basis of relative electronegativities. Except CoNbY$^\prime$Al (Y$^\prime$ = Fe,Mn) this empirical rule works well for all other quaternary compounds. Fixing the position of Al at $4a$ site we find Y and Zr having electronegativities lower than Co, Mn,Fe in CoX$^\prime$Y$^\prime$Al occupy $4b$ site leading CoYMnAl(CoYFeAl) and CoZrMnAl(CoZrFeAl) to crystallise in T$_{\text{I}}$ validating the empirical rule. Similarly for other  CoX$^\prime$Y$^\prime$Al (X$^\prime$ = Tc, Ru, Rh, Pd and Ag , Y$^\prime$=Mn, Fe) compounds, fixing the position of Z atom at $4a$ implies that the least electronegative one of the remaining three transition element atoms will occupy the $4b$ site producing the structure Type-II (Fig \ref{Crystal1}(b)). It is also to be noted that irrespective of the electronegativitiy of the $sp$-elements at the $4a$ site, $4b$ site is always occupied by the least electronegative of the transition metal elements.  The closeness in electronegativities of Nb and Mn (Co and Fe) (The electronegativities of Nb,Mn,Fe and Co are 1.6,1.55,1.83 and 1.88 respectively), might explain the reason behind CoNbY$^\prime$Al (Y$^\prime$ = Mn, Fe) being the exception to the empirical rule. Another exception is that of CoMoY$^\prime$Al. In t his case, Al being the least electronegative of all four atoms $4b$ site is occupied by the highest electronegative element Mo. Thus, the octahedral  sites are occupied by the main group element and the least electronegative of the remaining three making a rocksalt sub-lattice, while the tetrahedral sites are fixed with other two transition metal elements, leading to a structure type T$_{\text{I}}$. The energy differences between different structures with respect to ground state-type ($\Delta E$, $\Delta E^{\prime}$) are not very small indicating the ground state structure types, given in Table \ref{table1} are strongly preferred, unlike the cases of CoAgX$^{\prime}$Si compounds where the deviations from the empirical rule was explained in terms of closeness of energies in T$_{\text{I}}$ and T$_{\text{II}}$ \cite{NSRP}.  A comparison with CoX$^{\prime}$Y$^{\prime}$Si compounds point out another difference: the Mo-compounds crystallised in T$_{\text{II}}$ as expected from the empirical rules. The difference can be explained by the fact that Si has a electronegativity of 1.9, much higher than Al and certainly not the least among the four constituents.

The quaternary Heusler XX$^\prime$Y$^\prime$Z can be considered as a daughter compound derived from  the parent ternary compounds X$_2$Y$^\prime$Z \cite{DaiJAP09,CoFeMnSi_aftab,DrewsJLCM86}. Depending on the choice of parent ternary alloys there exist a number of possible combinations that yield quaternary Heuslers, at least theoretically. In this work we only considered X$_2$X$^\prime$Z and Y$_2^\prime$X$^\prime$Z as the parent compounds. For example, CoX$^{\prime}$Y$^{\prime}$Al can be considered to be derived from parents Co$_2$X$^{\prime}$Al and Y$^{\prime}_2$X$^{\prime}$Al. Considering T$_{\text{I}}$ as  equivalent to that of the regular Heusler structure  and T$_{\text{II}}$ as that of the ''inverse'' Heusler structure of ternary compounds we find that except CoTcMnAl,CoTcFeAl, CoMoMnSi, CoMoFeSi, CoPdMnSi, CoPdFeSi and CoAgMnSi all the CoX$^{\prime}$Y$^{\prime}$Z (Z=Al,Si) compounds follow the same structure type preferred by parent ternary alloys \cite{NSRP,srikrishna1}.Co$_2$TcAl (Mn$_2$TcAl) prefers T$_{\text{I}}$(T$_{\text{II}}$) as ground state \cite{srikrishna1} implying that the daughter compound can assume any of the two types depending on the energy cost. This is the case for other such deviant daughters as well. 

The calculated lattice constants agree well with the available results for select systems.The variations in the lattice constants across compounds are consistent with that in the atomic radii of X$^\prime$. Comparing the results in this work with those of CoX$^\prime$Y$^\prime$Si series, we find that change in lattice constant is in accordance with the changes in the size of the main group element.
 The following expression was used to calculate the formation energy \\
 
 $E_f = E_{CoX^{\prime}Y^{\prime}Al} - (E_{Co} + E_{X^\prime} + E_{Y^\prime} + E_{Al}) \quad\quad (i)$ \\
 
 Where $E_{CoX^{\prime}Y^{\prime}Al}$ is the total energies per formula unit of the CoX$^{\prime}$Y$^\prime$Al and $E_{Co}$, $E_{X{^\prime}}$, $E_{Y{^\prime}}$ and $E_{Al}$  are the total energies of the bulk Co, X$^\prime$,  Y$^\prime$ and Al respectively in their ground state structures. From Table \ref{table1},  we  see that  for all Quaternary compounds formation energies are negative, hence thermodynamically stable and are probable to form.

\begin{table*}%[t]
\caption{\label{table1}Calculated lattice constants, formation energies,
magnetic moments and energy differences between  possible structures  of quaternary
CoX$^{\prime}$Y$^{\prime}$Al compounds. The energy differences between Type-II (T$_{II}$) and
Type-I(T$_{I}$) structures are given by $\Delta$E = E$_\text{T$_\text{II}$}$ - E$_\text{T$_\text{I}$}$ .
A positive $\Delta$E implies T$_{I}$ as the ground state. The energy differences between
Type-III(T$_{III}$) and Type-I or Type-II structures are given by  $\Delta$E$^{\prime}$ = E$_\text{T$_\text{III}$}$ - E$_\text{T$_\text{I}$/T$_\text{II}$}$. For the systems which  crystallize in Type-II(T$_\text{II}$),  $\Delta$E$^{\prime}$ = E$_\text{T$_\text{III}$}$ - E$_\text{T$_\text{II}$}$;  for the systems which crystallize in Type-I(T$_\text{I}$), $\Delta$E$^{\prime}$ = E$_\text{T$_\text{III}$}$  - E$_\text{T$_\text{I}$}$.}
%\centering
%\resizebox{\linewidth}{!}{
\begin{tabular}
%{ l@{\hspace{0.5cm}} l@{\hspace{0.5cm}} c@{\hspace{0.5cm}} c@{\hspace{0.5cm}} c@{\hspace{0.5cm}} c@{\hspace{0.5cm}} l }
{ l@{\hspace{0.5cm}} l@{\hspace{0.5cm}} c@{\hspace{0.5cm}} c@{\hspace{0.5cm}} c@{\hspace{0.5cm}} c@{\hspace{0.5cm}} c@{\hspace{0.5cm}} c@{\hspace{0.5cm}} l }

%{ c c c c c c c c|c}
\hline\hline
Systems & Lattice  & Formation  & $\Delta E$ & $\Delta E^{\prime}$ &    Structure & M   \\
(Type-I)& constant(\AA) & energy(eV) & (eV/f.u.)        & (eV/f.u.)           &   type     & ($\mu_{B}/f.u.$)  \\ [0.5ex]
\hline
CoYMnAl & 6.41 (6.412\cite{coymnal_rahmoune2016}) & -2.01 & 0.70 & 0.35 &  T$_\text{I}$ & 4.00 (4\cite{coymnal_rahmoune2016})\\
CoZrMnAl & 6.08 (6.0746\cite{CoZrMnAl2018}) & -2.87 & 0.95 & 0.49 &  T$_\text{I}$ & 0.98 (1.0030\cite{CoZrMnAl2018})\\
CoNbMnAl & 5.94 & -2.41 & 1.19 & 1.09 &  T$_\text{I}$ & 0.00\\
CoMoMnAl & 5.87 & -1.77 & 0.34 & 0.88 &  T$_\text{I}$ & 0.91\\
CoTcMnAl & 5.85 & -2.94 & -0.47 & 0.92 &  T$_\text{II}$ & 2.11\\
CoRuMnAl & 5.85 & -3.42 & -1.05 & 0.73 &  T$_\text{II}$ & 3.02\\
CoRhMnAl & 5.88 (5.8924\cite{CoRhMnZJAC15}) & -2.98 & -1.34 & 0.80 &  T$_\text{II}$ & 4.05(4.04\cite{CoRhMnZJAC15})\\
CoPdMnAl & 5.97 & -2.68 & -1.09 & 0.73 & T$_\text{II}$ & 5.00\\
CoAgMnAl & 6.08 & -1.07 & -0.42 & 0.40 &  T$_\text{II}$ & 4.59\\

\hline
CoYFeAl & 6.22 & -1.76 & 0.74 & 0.67 & T$_\text{I}$ & 0.86\\
CoZrFeAl & 6.04 & -3.23 & 1.45 & 1.40 & T$_\text{I}$ & 0.00\\
CoNbFeAl & 5.93 (5.9482\cite{CoNbFeAl2018half}) & -2.36 & 1.15 & 1.42 & T$_\text{I}$ & 1.00 (1.00\cite{CoNbFeAl2018half})\\
CoMoFeAl & 5.87 & -1.63 & 0.50 & 1.04 &  T$_\text{I}$ & 1.93\\
CoTcFeAl & 5.86 & -2.33 & -0.12 & 0.60 &  T$_\text{II}$ & 3.05\\
CoRuFeAl & 5.85 & -3.09 & -0.74 & 0.62 &  T$_\text{II}$ & 4.20\\
CoRhFeAl & 5.88 & -2.60 & -1.12 & 0.60 &  T$_\text{II}$ & 4.86\\
CoPdFeAl & 5.94 & -2.08 & -0.79 & 0.29 &  T$_\text{II}$ & 4.41\\
CoAgFeAl & 6.02 & -0.64 & -0.37 & 0.10 &  T$_\text{II}$ & 3.82\\ [1ex]
\hline\hline
\end{tabular}%}
\end{table*}

%%%%%%%%%%%%%%%%%%%%%%%%%%%%%%%%%%%%%%%%%%%%%%%%%%%%%%%%%%%%%%%%%%%%%%%%%%%%%%%%%%%%%%%%%%%%%%%%%%

%%%%%%%%%%%%%%%%%%%%%%%%%%%%%%%%%%%%%%%%%%%%%%%%%%%%%%%%%%%%%%%%%%%%%%%%%%%%%%%%%%%%%%%%%%%%%
%========================= S-P rule ==================================
%clearpage
\subsection{The variations of magnetic moments across compounds and possibility of half-metallicity}

It is well known that half-metallic Heusler alloys possess integer spin moment and follow the  Slater-Pauling rule that provides an empirical relation between the total moment and N$_{\text{V}}$, the total number of valance electrons per formula unit \cite{SlaterPRB36,PaulingPRB38,Galanakis_PRB13,Galanakis2016}. The Slater-Pauling rule states that the total moment $M$ is linearly proportional to the total number of valance electron counts by M =$|$N$_{\text{V}}$ - 18$|$ or M =$|$N$_{\text{V}}$ - 24$|$ or M =$|$N$_\text{V}$ - 28$|$ depending on whether the  the Fermi-energy is placed in an energy  gap after 9, 12 or 14 states respectively \cite{GalanakisPRB02,Galanakis_PRB13,GalanakisJAP13_113}.The variations in the total spin-magnetic moment per formula unit (M) with N$_{\text{V}}$ for  the compounds considered in this work are shown in Fig \ref{sp-rule-XX'Y'Al}.  Most of the quaternary Heusler compounds in our study are found to follow M =$|$N$_{\text{V}}$ -24$|$ rule  closely except for the cases when X$^{\prime}$ are late transition metal elements, CoAgMnAl, CoPdFeAl, CoAgFeAl which deviate substantially. The compound CoMnYAl has total moment 4 $\mu_B/f.u.$ with N$_{\text{V}}$ = 22, thus follows M =$|$N$_{\text{V}}$ -18$|$ rule. Qualitatively, this is similar to the trends for CoX$^{\prime}$Y$^{\prime}$Si series \cite{NSRP}.
However, only seven compounds from the present study, CoYMnAl, CoZrMnAl, CoRuMnAl, CoRhFeAl, CoPdMnAl, CoNbFeAl, and CoTcFeAl have either integer or near-integer magnetic moments and thus, are potential half-metals. In the following, their electronic structures will be critically examined to ascertain the half-metallic behaviour.

%%%%%%%%%%%%%%%%%%%%%%%%%%%%%%%%%%%%%%%%%%%%% For X2X'Al and XX'Y'Al %%%%%%%%%%%%%%%%%%%%%%%%%%%%%%
	\begin{figure}[H]
            \centerline{\hfill
	    \psfig{file=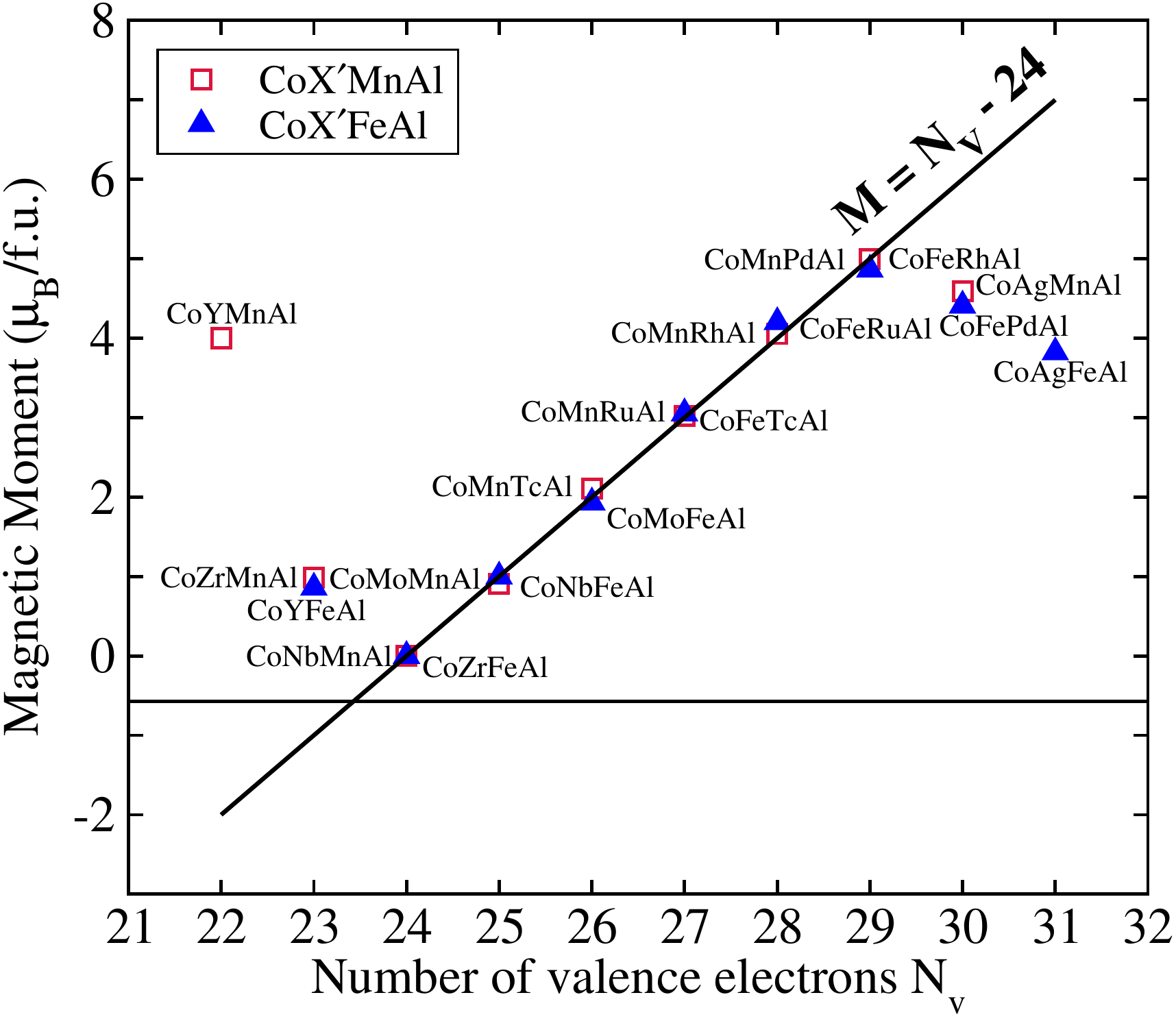,width=0.5\textwidth}\hfill}
            \caption{Total magnetic moments versus the total number of 
             valence electrons N$_{\text V}$ for CoX$^{\prime}$Y$^\prime$Al compounds. The Slater-Pauling
             M=N$_{\text V}$-24  line is drawn as a guide.  }
\label{sp-rule-XX'Y'Al}
          \end{figure}

%%%%%%%%%%%%%%%%%%%%%%%%%%%%%%%%%%%%%%%%%%%% For X2X'Si %%%%%%%%%%%%%%%%%%%%%%%%%%%%%%

%========================= Electronic Structure ==================================
%\clearpage
\subsection{Electronic Structure}
%====================================================================================================
The origin of half-metallic behaviour in Heusler compounds with 3$d$ transition metal atoms have been well explained \cite{GalanakisPRB02,bouckaert1936lp}. In References \onlinecite{NSRP,srikrishna1} it has been shown that half-metallic behaviour in ternary and quaternary Heuslers with both 3$d$ and 4$d$ transition metals together, can be explained the same way. In References \onlinecite{NSRP,srikrishna1} it has been demonstrated that the positions of the bonding e$_{g}$, t$_{2g}$ and non-bonding t$_{u}$, e$_{u}$ orbitals, arising out of the hybridisations of the transition metal atoms, with respect to the Fermi levels are responsible for the half-metallicity or lack of it in these compounds. The positions of the $s$ and $p$ bands of the main group elements are also important as these states accommodate charges from the $d$ states and thus play a role in determining the positions of the $d$ bands. In this sub-section we present the spin-polarised total and atom projected densities of states for all  CoX$^\prime$Y$^\prime$Al compounds in their respective ground state structures, determine the compounds that are half-metals and compare them with the compounds in the Si series \cite{NSRP} to obtain a generalised picture across series of compounds.

The densities of states of all compounds are presented in Fig \ref{DOSCoX'Y'Al-t1} and Fig \ref{DOSCoX'Y'Al-t2}. 
 Fig \ref{DOSCoX'Y'Al-t1}.(a) and Fig \ref{DOSCoX'Y'Al-t1}.(b)  show the electronic structures of the compounds with structure type T$_\text{I}$. We find that CoYMnAl and CoNbFeAl are the two half-metals with a semiconducting gap in the spin down band, resulting in 100$\%$ spin polarisation. CoZrMnAl and CoMoFeAl are another pair of compounds with high spin polarisations (more than 90$\%$). Their electronic structures too have distinct features of a near gap in one of the spin bands. In case of CoYMnAl, the gap originates due to separation of t$_{2g}$ (due to hybridisation of all transition metal $d$ orbitals) and t$_{u}$ states (due to Co and Mn states). The pattern of hybridisation is same as CoYMnSi \cite{NSRP}. However, CoYMnSi is not a half-metal as the t$_{u}$ states were below the Fermi level. This presumably occurred as the the deeper lying Si states (compared to Al) pulled the $d$ states downward in energy. In case of CoNbFeAl, the gap is created due to the separations of t$_{u}$ and e${_u}$ states. CoZrFeAl, having exactly 24 electrons, have equal distributions among the spin channels, resulting in a zero moment and opening of gaps in both spin channels. The extra electron in CoNbFeAl is accommodated in the spin up band, thus retaining the semiconducting gap in the spin down band. One more electron in CoMoFeAl, fills the spin up states almost completely. However, the t$_{2g}$ bands from the occupied part of the spectrum in the spin down channel move closer to the Fermi level producing small number of states there and reducing the spin polarisation to 96$\%$. This material, therefore, can be categorised as a "near half-metal" with high spin polarisation. Among the other members of CoX$^{\prime}$MnAl, we do not find any more half-metal due to the nature of band filling. CoZrMnAl has features very similar to CoZrFeAl in it's spin up densities of states. The highlight is a small near-gap bordered with t$_{u}$ and e$_{u}$ non-bonding states, while the extra electron as compared to CoYMnAl fills the spin down band. As a result, the system is a near half-metal with a spin polarisation of 95$\%$. CoNbMnAl has exactly 24 electrons, distributed evenly between two spin bands rendering this compound a non-magnetic semiconductor like CoZrFeAl. In fact the electronic structure of CoNbMnAl changes drastically with respect to CoZrMnAl. The densities of states of CoMoMnAl is quite similar to that of CoMoFeAl with a near-gap in the spin down band bordered by t$_{u}$ and e$_{u}$ hybrids. However, the position of the $e$ states of CoMoMnAl is lower in energy in comparison to those of CoMoFeAl, producing states at the Fermi level thereby lowering of the spin polarisation to 86$\%$.  
 
 Among the compounds with structure type T$_{\text{II}}$, we do not find any half-metal in CoX$^{\prime}$Y$^{\prime}$Al series. Only CoRuMnAl can be considered a near half-metal as the spin polarisation is 94$\%$. Qualitatively this is distinctly different from the Si-series quaternary compounds where two half-metals were found \cite{NSRP}. Among the CoX$^{\prime}$MnAl compounds with structure T$_{\text{II}}$, we find that in cases of four out of five compounds, there is indeed a gap in the spin down band. However, the gap does not extend through the Fermi level for any of the compounds. Once again this is due to the positions of the $t$ bands, primarily, with respect to the Fermi levels. A comparison with Si-series compounds \cite{NSRP} reveals that the $t$ bands in Si-series compounds were lying deeper in the spectrum, ostensibly due to the presence of deep lying Si states pulling the $d$ bands of transition metals towards lower energies. In case of CoX$^{\prime}$FeAl compounds, the extra electrons as one goes from T$_{\text{I}}$ CoMoFeAl to CoTcFeAl and onwards, are accommodated in the spin down $e$ bands, thus, pulling them closer to Fermi levels and destroying any possibility of having a half-metallic gap. Similar was the situation for CoX$^{\prime}$FeSi series with structure type T$_{\text{II}}$. The band structures of three compounds CoYMnAl, CoZrMnAl and CoRuMnAl are shown in Fig 1, supplementary material, to corroborate the inferences drawn from the densities of states. 
 
 A comparison between CoX$^{\prime}$Y$^{\prime}$Al and CoX$^{\prime}$Y$^{\prime}$Si series show one interesting pattern: the densities of states for compounds with same N$_{\text{V}}$ have striking resemblances. CoYMnSi and CoZrMnAl have similar features in the densities of states in both spin channels. So are CoZrMnSi-CoNbMnAl, CoMoMnAl-CoNbMnSi, CoYFeSi-CoZrFeAl,CoZrFeSi-CoNbFeAl,CoNbFeSi-CoMoFeAl. There are resemblances too in the spin down bands of CoMoMnSi-CoRuMnAl, CoTcMnSi-CoRhMnAl, CoRuMnSi-CoPdMnAl and CoRhMnSi-CoAgMnAl pairs, all with structure type T$_{\text{II}}$. This can be easily understood by considering how the bands in one of the spin channels are filled continuously, depending upon N$_{\text{V}}$, across the series.
 
 \begin{figure}[h!]
\centering
\subfigure[]{\includegraphics[scale=0.22]{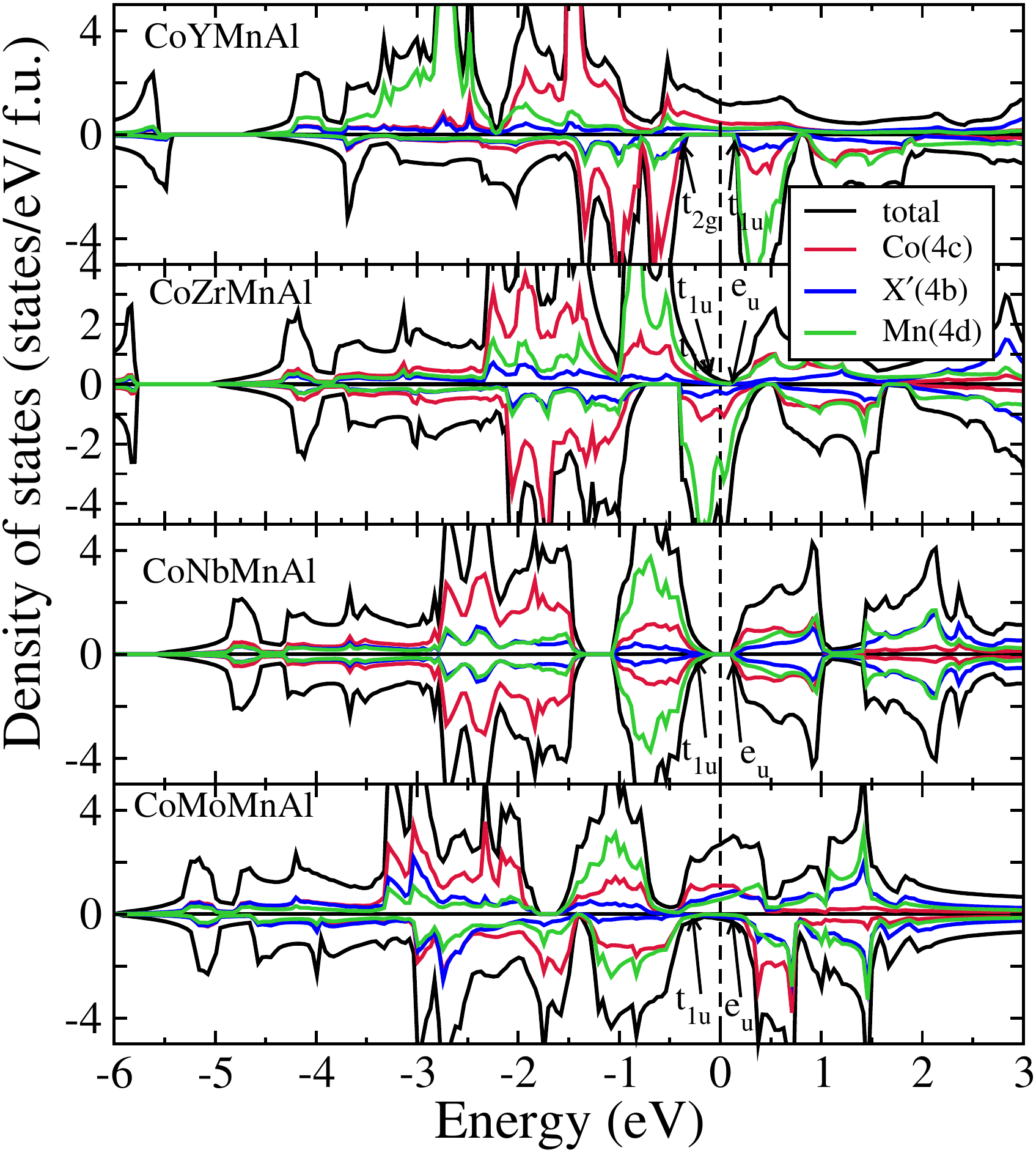}}
\subfigure[]{\includegraphics[scale=0.22]{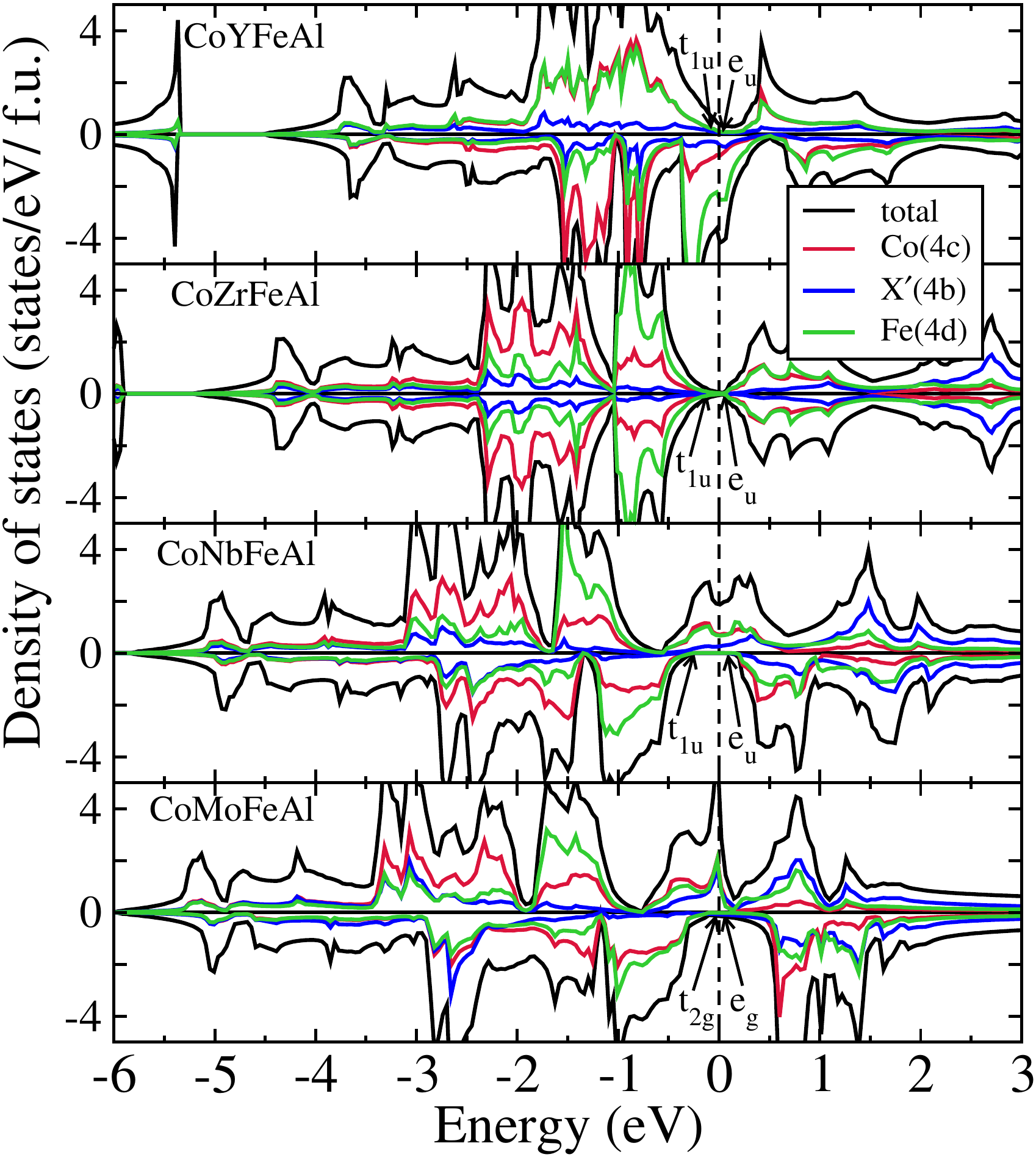}}
\caption{Spin polarised total and atom-projected densities of states for ({\bf a}) CoX$^\prime$MnAl and ({\bf b}) CoX$^\prime$FeAl (X$^\prime$ = Y, Zr, Nb, Mo) compounds. The ground states of these compounds are T$_{\text{I}}$. }
\label{DOSCoX'Y'Al-t1}
\end{figure}

\begin{figure}[h!]
\centering
\subfigure[]{\includegraphics[scale=0.22]{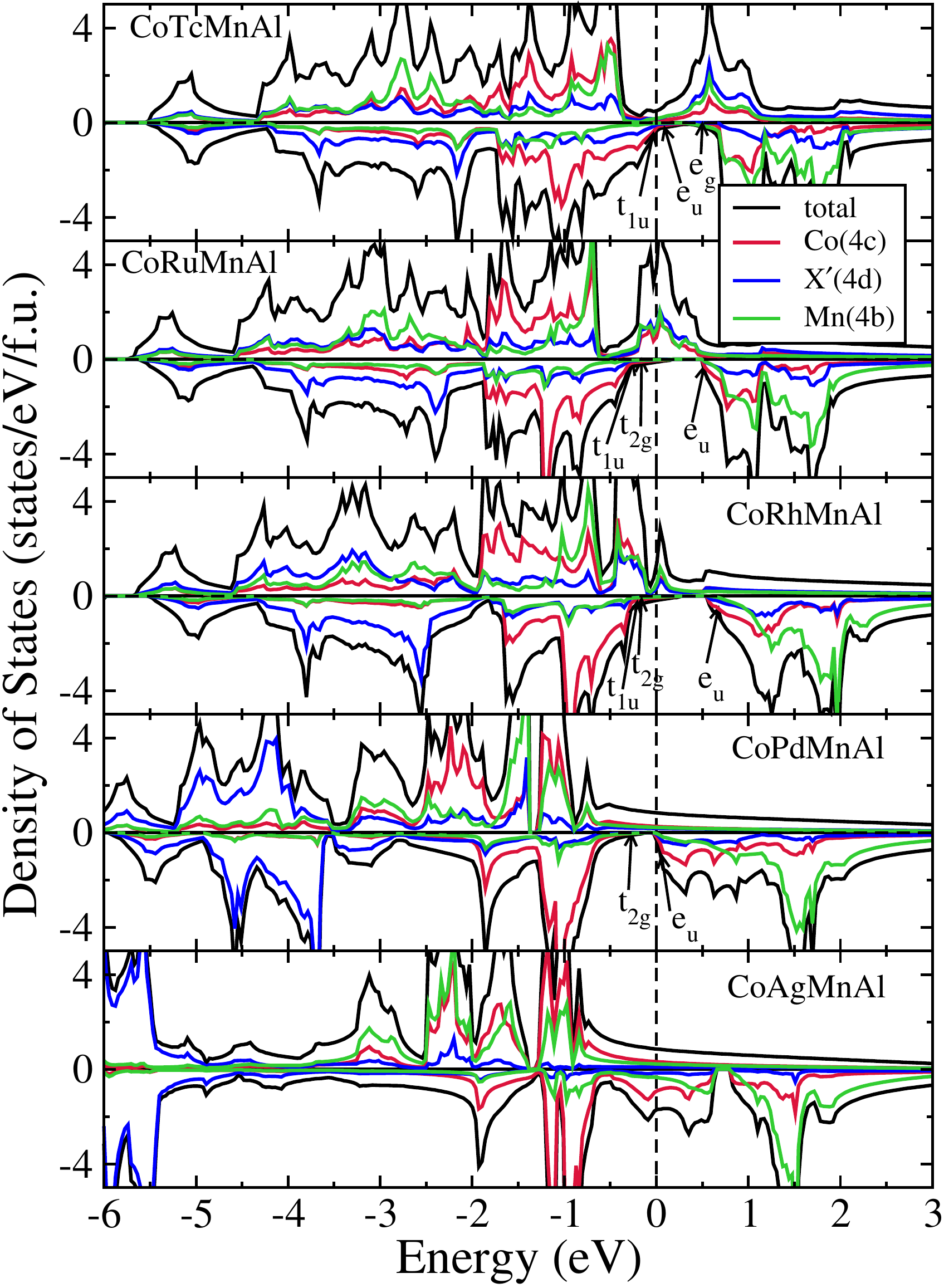}}
\subfigure[]{\includegraphics[scale=0.22]{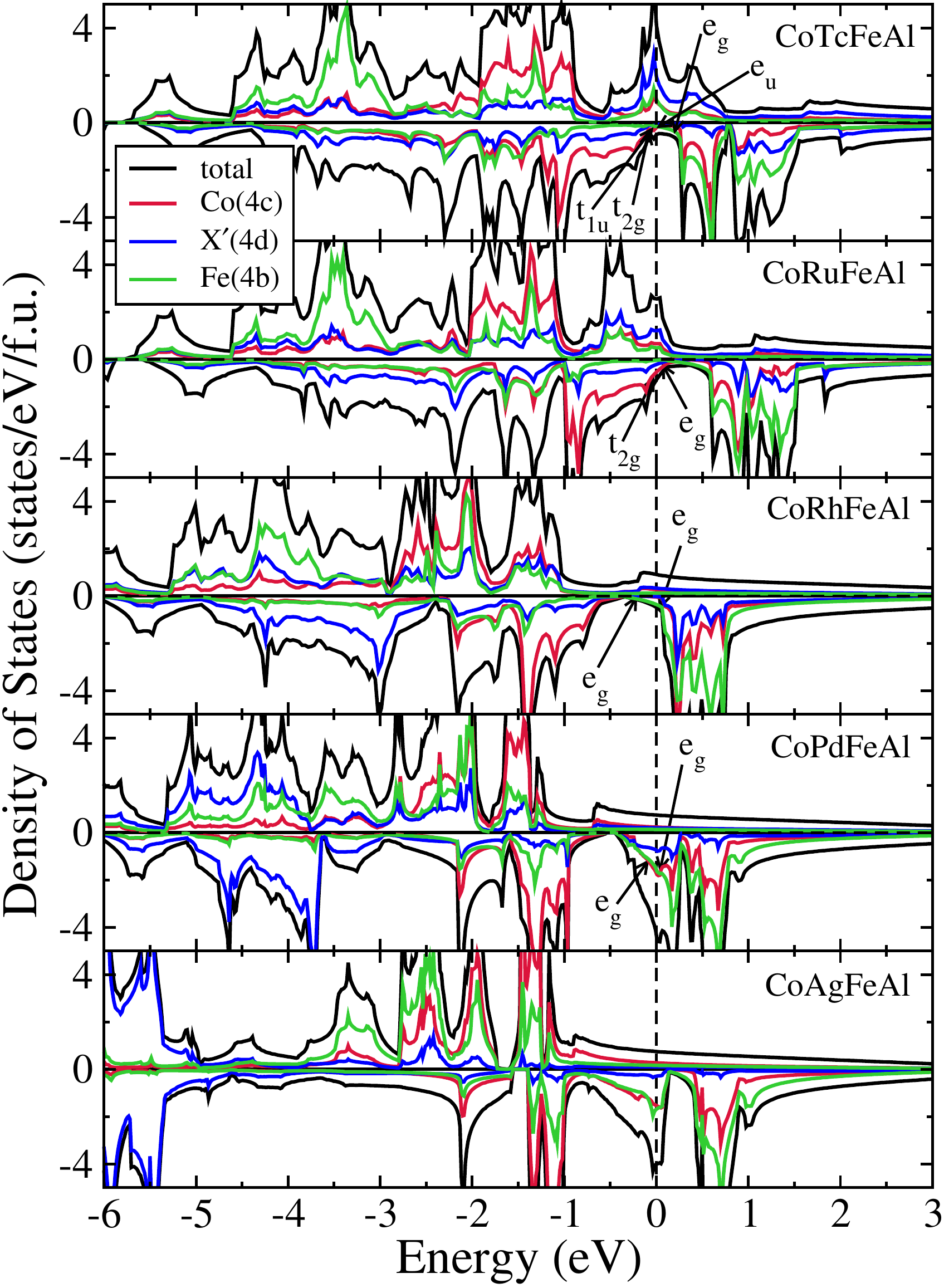}}
\caption{Spin polarized total and atom-projected densities of states for ({\bf a}) CoX$^\prime$MnAl and ({\bf b}) CoX$^\prime$FeAl (X$^\prime$ = Tc, Ru, Rh, Pd, Ag) compounds. The ground states of these compounds are Type-II. }
\label{DOSCoX'Y'Al-t2}
\end{figure}

\subsection{Variations in the atomic moments with N$_{\text{V}}$ and structure type}
 In Tables \ref{table2} and  \ref{table3} we present the total and site projected magnetic moments of the compounds considered in this work. Analysing the trends as a function of N$_{\text{V}}$ and structure type, we find near identical behaviour with CoX$^{\prime}$Y$^{\prime}$Si series \cite{NSRP}. Across the series and structures, the total moment continuously increases with N$_{\text{V}}$ for compounds with N$_{\text{V}} >$ 24 and upto 29. This is understood from the hybridisation and band filling picture discussed in the previous sub-section. We find that the continuous building up of the moment is due to building up of Co and Y$^{\prime}$ moments. This is more prominent for compounds in structure type T$_{\text{II}}$ where the moments on the 3$d$ elements build up due to they being nearest neighbours leading to large exchange splitting primarily on Y$^{\prime}$ elements. The changes in Fe moments with N$_{\text{V}}$ are smaller compared to Mn moments, implying Fe moments are more localised. The moments on the 4$d$ compounds are small. They gain moment due to polarising effect of strong exchange fields of 3$d$ elements. For compounds with N$_{\text{V}} >$ 29, the total moments start decreasing with increasing N$_{\text{V}}$. The corresponding compounds are CoPdFeAl, CoAgFeAl and CoAgMnAl. The electronic structure reveals that the Pd states are deep in energy and both spin bands are nearly filled. Since it has a filled $d$ shell, this is only natural that there will be a relatively weak exchange field associated with Pd. As a result, the moment on Pd drastically reduces compared to the X$^{\prime}$ element in the preceding compound in the series. This brings the overall moment down. Ag has $s$ states in it's valence band, which are further deep in energy leaving no scope of any spin polarisation of the X$^{\prime}$ component. Therefore, even though the Co and Y$^{\prime}$ moments do not change significantly, the moments of CoAgFeAl and CoAgMnAl decrease with respect to the preceding compounds in their respective series. 
 
 The exchange splittings in compounds with structure type T$_{\text{I}}$ are smaller, an artefact of the neighbourhood around the 3$d$ components. The exception is CoYMnAl where Mn has a large moment. A significantly large lattice constant in this compound is a hindrance for Mn to hybridise with the other components leading to it retaining it's local moment in the atomic state. The densities of states of this compound is consistent with this assessment. One can see that in the spin up channel, the Mn states are deep lying and are well separated from the Co states, along with the fact that Mn down band is nearly empty. 
 \begin{table}[H]
\caption{\label{table2}Total and atomic magnetic moment of CoX$^{\prime}$MnAl systems in $\mu_{B}/f.u.$.  N$_{\text{V}}$ is the number of valence electrons of the systems. M is the total magnetisation.}
\vspace{2 mm}
\centering
%\resizebox{\linewidth}{!}{
\begin{tabular}{ c c c c c c c c c c}
\hline\hline
Systems(T$_{\text{I}}$) &  N$_{\text{V}}$ & M     & M$_{\text{Co}}$ & M$_{\text{Mn}}$ & M$_{\text{X$^{\prime}$}}$ & M$_{\text{Al}}$  & P($\%$)  \\ [0.1ex]
\hline
CoYMnAl 	 & 22 	 & 4.00 	 & 0.80 	 & 3.41 	 & -0.13 	 & -0.04 	 & 100 \\
CoZrMnAl 	 & 23 	 & 0.98 	 & 0.10 	 & 0.94 	 & -0.07 	 & -0.01 	 & 95 \\
CoNbMnAl 	 & 24 	 & 0.00 	 & -0.00 	 & 0.00 	 & -0.00 	 & -0.00 	 & 0 \\
CoMoMnAl 	 & 25 	 & 0.91 	 & 0.82 	 & -0.09 	 & 0.22 	 & -0.01 	 & 86 \\[1ex]
\hline
Systems(T$_{\text{II}}$)& N$_{\text{V}}$  & M     & M$_{\text{Co}}$ & M$_{\text{X$^{\prime}$}}$ & M$_{\text{Mn}}$ & M$_{\text{Al}}$  & P($\%$)   \\ [0.1ex]
\hline
CoTcMnAl 	 & 26 	 & 2.11 	 & 0.57 	 & -0.62 	 & 2.21 	 & -0.02 	 & 20 \\
CoRuMnAl 	 & 27 	 & 3.02 	 & 0.53 	 & -0.13 	 & 2.66 	 & -0.01 	 & 94 \\
CoRhMnAl 	 & 28 	 & 4.05 	 & 0.84 	 & 0.25 	 & 3.07 	 & -0.03 	 & 77 \\
CoPdMnAl 	 & 29 	 & 5.00 	 & 1.27 	 & 0.25 	 & 3.48 	 & -0.03 	 & 39 \\
CoAgMnAl 	 & 30 	 & 4.59 	 & 1.07 	 & 0.03 	 & 3.48 	 & -0.03 	 & 33 \\ [1ex]
%%%%%%%%%%%%%%%%%%%%%%%%%%%%%%%%%%%%%%%%%%%%%%%%% Copied from Mn_pol_latex , created in POLARIZATION folder and copied here ##################

\hline\hline
\end{tabular}%}
\end{table}
%==========================================================================================

\begin{table}[H]
\caption{\label{table3}Total and atomic magnetic moments of Co$\text{X}^\prime$FeAl systems in $\mu_{B}/f.u.$.  N$_\text{V}$ is the number of valence electron of the systems. M is the total magnetisation.}
\vspace{2 mm}
\centering
%\resizebox{\linewidth}{!}{
\begin{tabular}{ c c c c c c c c c c}
\hline\hline
Systems(T$_\text{I}$) &  N$_\text{V}$ & M & M$_\text{Co}$ & M$_\text{Fe}$ & M$_\text{X$^{\prime}$}$ & M$_\text{Al}$ & P(\%)  \\ [0.5ex]
\hline
CoYFeAl 	 & 23 	 & 0.86 	 & 0.04 	 & 0.93 	 & -0.04 	 & -0.02 	 & 84 \\
CoZrFeAl 	 & 24 	 & 0.00 	 & 0.00 	 & -0.00 	 & 0.00 	 & 0.00 	 & 0 \\
CoNbFeAl 	 & 25 	 & 1.00 	 & 0.62 	 & 0.56 	 & -0.09 	 & -0.01 	 & 100 \\
CoMoFeAl 	 & 26 	 & 1.93 	 & 1.14 	 & 0.77 	 & 0.10 	 & -0.01 	 & 96 \\ [1ex]
Systems(T$_\text{II}$) &  N$_\text{V}$ & M & M$_\text{Co}$ & M$_\text{X$^{\prime}$}$ & M$_\text{Fe}$ & M$_\text{Al}$  & P(\%)\\ [0.5ex]
\hline
CoTcFeAl 	 & 27 	 & 3.05 	 & 0.77 	 & -0.14 	 & 2.43 	 & 0.01 	 & 78 \\
CoRuFeAl 	 & 28 	 & 4.20 	 & 1.14 	 & 0.40 	 & 2.77 	 & -0.01 	 & 51 \\
CoRhFeAl 	 & 29 	 & 4.86 	 & 1.40 	 & 0.55 	 & 3.00 	 & -0.02 	 & 32 \\
CoPdFeAl 	 & 30 	 & 4.41 	 & 1.37 	 & 0.15 	 & 2.92 	 & -0.02 	 & 72 \\
CoAgFeAl 	 & 31 	 & 3.82 	 & 1.20 	 & -0.02 	 & 2.74 	 & -0.03 	 & 69 \\ [1ex]
\hline\hline
\end{tabular}%}
\end{table}

%%%%%%%%%% Copied from Fe_pol_latex , created in POLARIZATION folder and copied here ##################

\subsection{Variations in magnetic transition temperature T$_{\text{c}}$ across series}

The results on inter-atomic effective magnetic exchange coupling constants (J$^\text{eff}_{\mu\nu}=\sum_{j}J^{0j}_{\mu\nu};0$ fixed on sub-lattice $\mu$ and $j$ runs over sub-lattice $\nu$) and magnetic transition temperatures T$_{\text{c}}$ for the compounds considered in our study are presented in Figs \ref{curie-temp-cox'y'al}-\ref{exchange-cox'feal}. Fig \ref{curie-temp-cox'y'al} shows the variations of T$_{\text{c}}$ with different X$^\prime$ atoms {\it i.e.} with changes in N$_{\text{V}}$ for  CoX$^\prime$Y$^\prime$Al. The two half-metals CoYMnAl and CoNbFeAl have T$_{\text{c}}$ of 430K and 232K respectively. Our result for CoYMnAl agrees well with the value of 482K obtained by Rahmoune {\it et al}\cite{coymnal_rahmoune2016}. Looking at Fig \ref{curie-temp-cox'y'al} we find that qualitative variations in T$_{\text{c}}$ depends on the  ground state structure type of the compounds in the series implying that the arrangements of the atoms in crystallographic sites play a significant role in deciding T$_{\text{c}}$ as well. We find that in general, compounds with structure type T$_{\text{II}}$ have T$_{\text{c}}$ higher than those of compounds with structure type T$_{\text{I}}$. The highest T$_{\text{c}}$ for both series is obtained at N$_{\text{V}}$=29, a result exactly same as the quaternary Si-series \cite{NSRP}. Similar to CoX$^\prime$Y$^\prime$Si series\cite{NSRP} we find that, Co-Y$^\prime$ J$^\text{eff}$ are the dominant interactions  and the variations in T$_{\text{c}}$ closely follow the  variations in this particular effective exchange parameter. 
Other prominent exchange parameters are for Co-Co, Y$^\prime$-Y$^\prime$, Y$^\prime$-X$^\prime$ ones.  As was seen in Si-based compounds \cite{NSRP}, Co-Y$^\prime$ interactions for compounds in structure type T$_{\text{II}}$ are notably stronger than those of compounds in structure T$_{\text{I}}$. The reason for this is that, Co-Y$^\prime$ interaction is direct in T$_{\text{II}}$ whereas weak magnet X$^\prime$ is the mediator for Co and Y$^\prime$ interactions in T$_{\text{I}}$.  

Though qualitatively the variations of T$_{\text{c}}$ across series and structure types in CoX$^{\prime}$Y$^{\prime}$Al are exactly same as that in CoX$^{\prime}$Y$^{\prime}$Si compounds, there are quantitative differences. In general, CoX$^{\prime}$FeAl (CoX$^{\prime}$MnAl)have higher (lower)T$_{\text{c}}$ than CoX$^{\prime}$FeSi(CoX$^{\prime}$MnSi). The difference is prominent in compounds with T$_{\text{II}}$ structure type. In CoX$^{\prime}$FeAl compounds with structure T$_{\text{II}}$, we find that Co-Fe and Fe-X$^{\prime}$ interactions are stronger than those in CoX$^{\prime}$FeSi compounds. This along with more ferromagnetic Fe-Fe interactions in Al-series compounds lead to higher T$_{\text{c}}$ in them. In the CoX$^{\prime}$MnZ compounds, we find that in Si-series, the dominant Co-Mn interactions are stronger than that in Al-series; the other prominent interactions being of comparable strength. This, then, decides the quantitative differences in T$_{\text{c}}$ of the two series. 

Finally the reason behind T$_{\text{c}}$ being maximum at N$_\text{V}$ = 29 across the series irrespective of Y$^\prime$  and Z  element can be explained in terms of ``Exchange average''\cite{KublerPRB07}, exactly the way was done in Reference \onlinecite {NSRP}. 
Inspecting the densities of states for compounds with structure type T$_{\text{II}}$, we see that  lower the value of the number of states in spin down channel at Fermi level in the compounds with structure Type-II,  higher is the ``Exchange average'' and subsequently the T$_{\text{c}}$. For both Al and Si-series quaternary compounds we see that as N$_\text{V}$ increases, the number of states at the Fermi level keeps decreasing and that in turn increases the ``Exchange average'' and T$_{\text{c}}$  upto a particular N$_\text{V}$. The subsequent fall of T$_{\text{c}}$ can also be attributed to the fall in  ``Exchange average''  with increasing N$_\text{V}$ as the electronic structure shows a fall in the number of available  states at the Fermi level in spin down bands. Thus we conclude that the significant similarity in the qualitative trends in the variations of T$_{\text{c}}$ across Y$^{\prime}$ and Z series of Co and 4$d$ transition metal based quaternary Heuslers can be correlated to the similarities in the trends in the variations in their electronic structures which is dependent upon structure types and N$_{\text{V}}$.

%%%%%%%%%%%%%%%%%%%%%%%%%%%%%%%%%% Curie temperature and Exchange parameters of CoY'X'Al %%%%%%%%%%%%%%%%%%%%%%%%%%%%%%%%%

%%%%%%%%%%%%%%%%%%%%%%%%%%%%%%%%%%%%%%%%%%%%%%%%% CoX'Y'Al %%%%%%%%%%%%%%%%%%%%%%%%%%%%%%%%%%%%%%%%%%%%%%%
\begin{figure}[H]
\centerline{\hfill
\psfig{file=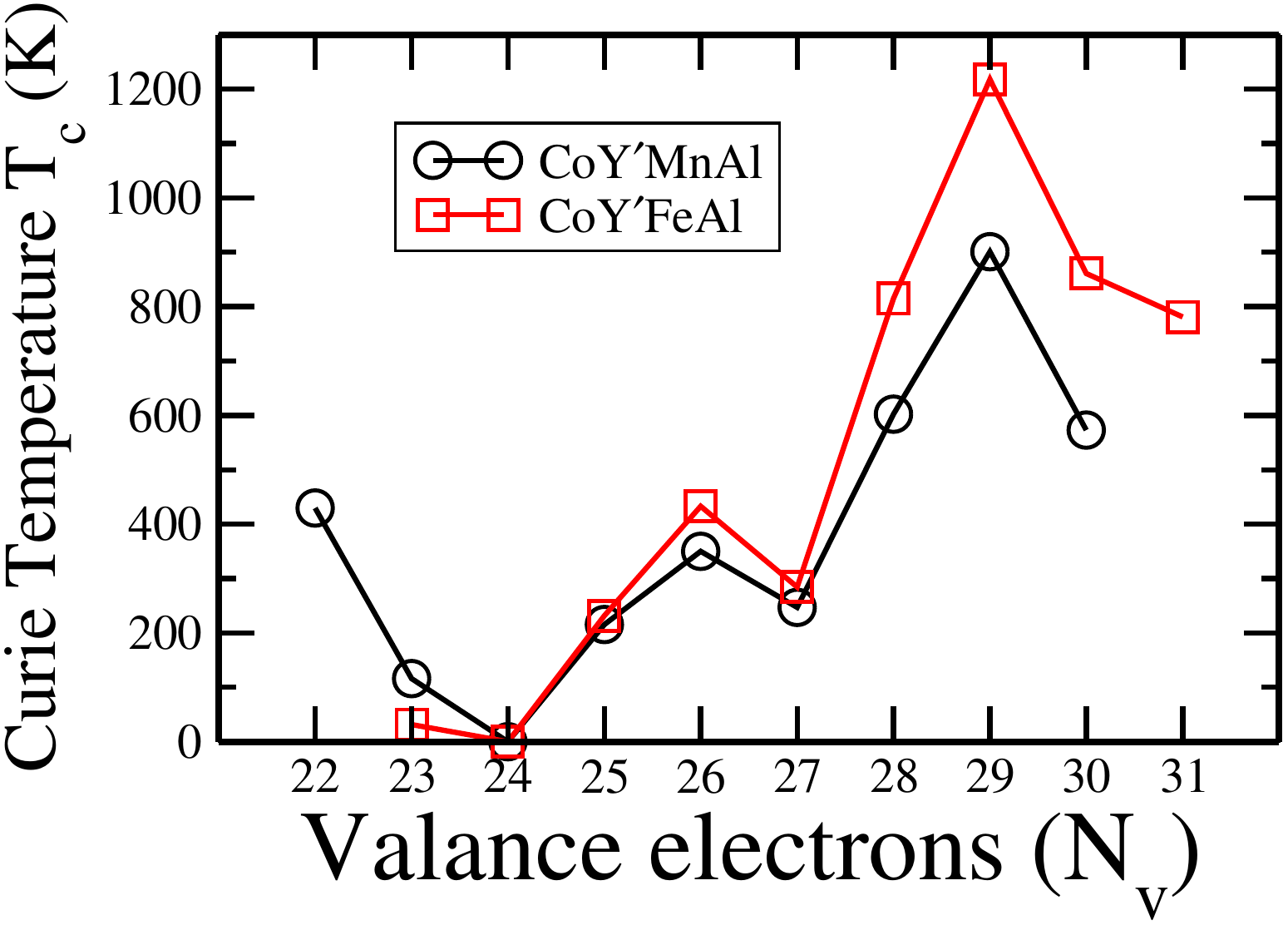,width=0.35\textwidth}\hfill}
 \caption{Calculated Curie temperature with total number of valence electron for CoY$^\prime$X$^\prime$Al series.}
\label{curie-temp-cox'y'al}
\end{figure}

\begin{figure}[H]
\centerline{\hfill
\psfig{file=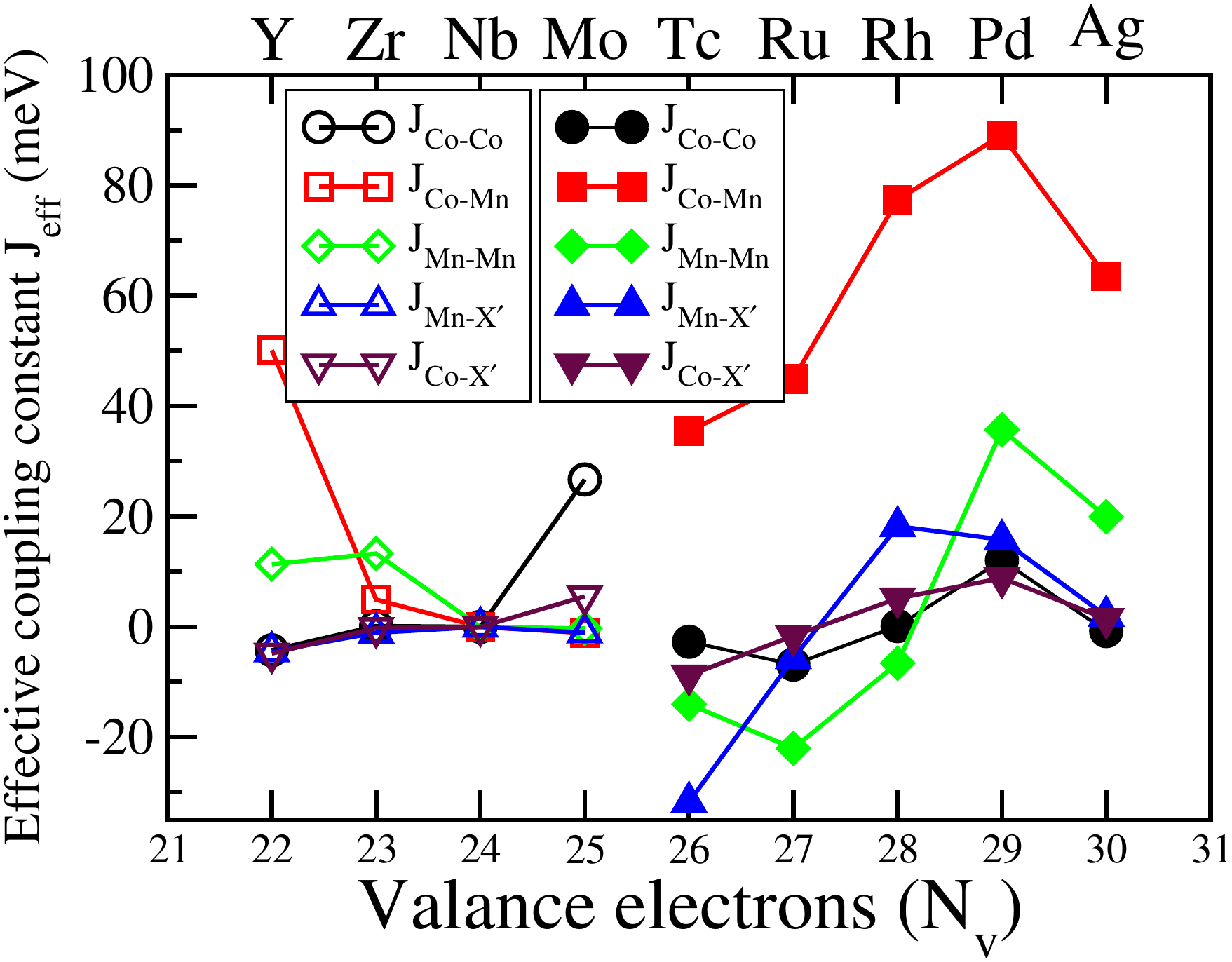,width=0.35\textwidth}\hfill}
\caption{Effective exchange coupling constant for Co$\text{X}^\prime$MnAl alloys. Open symbols: structure type T$_{\text{I}}$, Filled symbols: structure type T$_{\text{II}}$.}
\label{exchange-cox'mnal}
\end{figure}

\begin{figure}[H]
\centerline{\hfill
\psfig{file=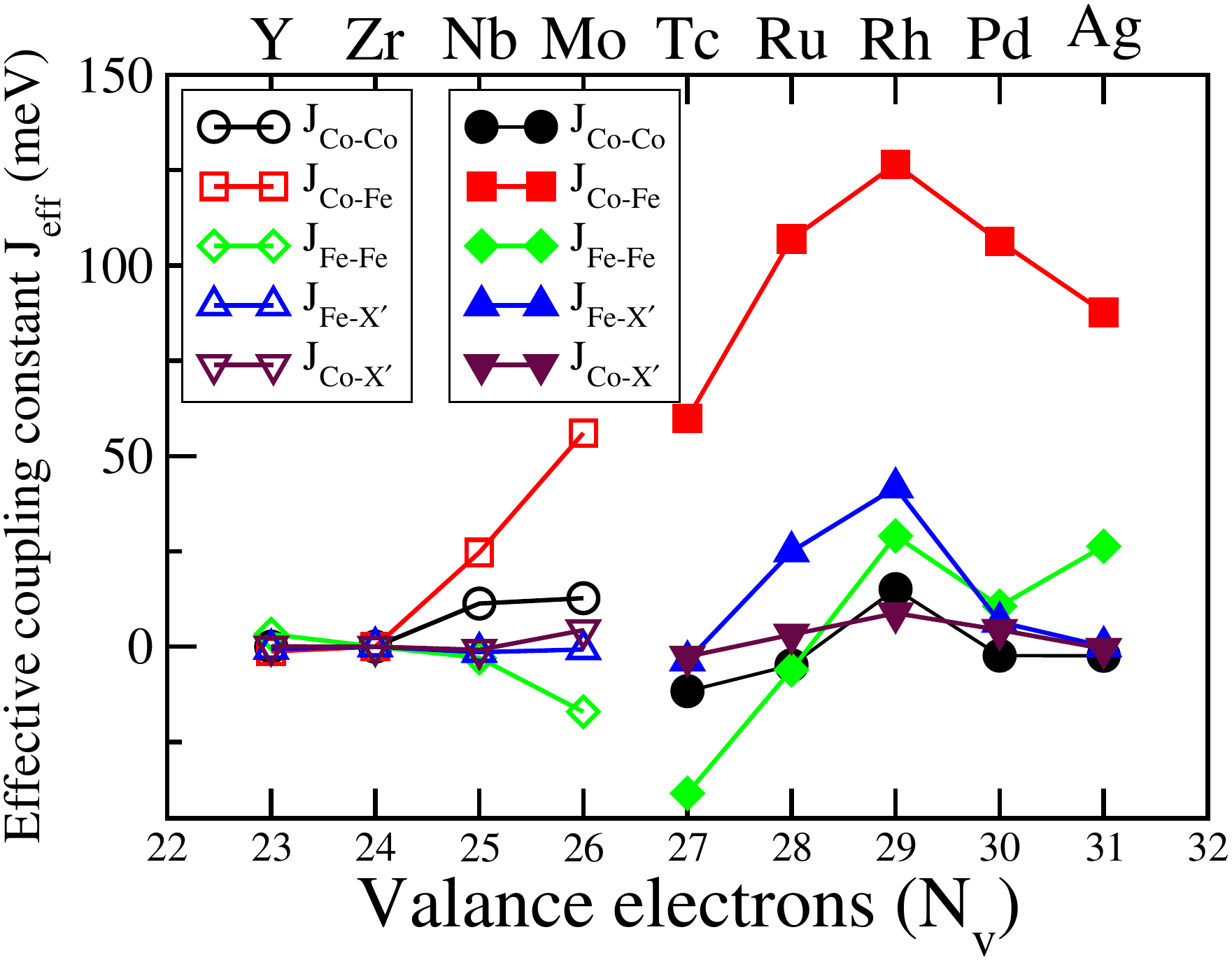,width=0.4\textwidth}\hfill}
\caption{Effective exchange coupling constant for Co$\text{X}^\prime$FeAl alloys. Open symbols: structure type T$_{\text{I}}$, Filled symbols: structure type T$_{\text{II}}$.}
\label{exchange-cox'feal}
\end{figure}
%%%%%%%%%%%%%%%%%%%%%%%%%%%%%%%%%%%%%%%%%%%%%%%%%%%%%%%%%%%%%%%%%%%%%%%%%%%%%%%%%%

\subsection{Qualitative comparison with quaternary Heuslers having all magnetic components 3$d$ transition metals }

Our results in this paper and in Reference \onlinecite{NSRP} provide significant insights into the roles of atomic arrangement, the number of valence electrons, and the hybridisations towards achieving half-metallic behaviour. In order to gain deeper insights and evolve a generalised picture of half-metallicity in quaternary Heusler compounds, we now make a qualitative comparison between the series where the only difference is that X$^{\prime}$ a 4$d$ element in one and a $3d$ element (one of Sc to Cu in the row of 3$d$ transition metals in the periodic table) in another. In Table I, supplementary material, we have compiled the available results on the quaternary compounds CoX$^{\prime}$Y$^{\prime}$Z with main group element either Si or Al, Y$^{\prime}$ either Mn or Fe and X$^{\prime}$ one of the 3$d$ transition metals. From the compiled results we find that the in the quaternary family with all magnetic components 3$d$ transition metals, there are significantly higher number of materials reported to be half-metals or even SGS \cite{GalanakisJAP13_113,XuEPL13,CoFeMnSi_aftab}. Since the results in the present work, that of Reference \onlinecite{NSRP} and \onlinecite{srikrishna1} on ternary and quaternary Heusler alloy series with one of the magnetic element a 4$d$ transition metal indicate that N$_{\text{V}}$ is a significant predictor of half-metallic behaviour, we first compare the electronic properties, in the context of half-metallic behaviour, of compounds across the families with same N$_{\text{V}}$. 
    
We first observe that there are four pairs of half-metals (with same N$_{\text{V}}$), one each from two families, one with X$^{\prime}$ a 3$d$ element and another with X$^{\prime}$ the isoelectronic 4$d$ element. They are CoNbFeAl-CoVFeAl, CoZrFeSi-CoTiFeSi, CoTcMnSi-Mn$_{2}$CoSi and CoRhMnSi-Co$_{2}$MnSi. Strictly speaking the X$^{\prime}$ a 3$d$ element contingent in the last two pairs is not a quaternary compound. The ground state structures of all four compounds in the first two pairs are T$_{\text{I}}$ while that of the third pair is T$_{\text{II}}$. In the fourth pair, CoRhMnSi has T$_{\text{II}}$ structure but Co$_{2}$MnSi crystallises in T$_{\text{I}}$. There are another five pairs of compounds having nearly identical behaviour in the context of half-metallicity. In the pair CoMoFeAl-CoCrFeAl, CoMoFeAl is a near half-metal while CoCrFeAl is a SGS; both compounds in the pair CoNbMnSi-CoVMnSi are near half-metals. In the pairs CoZrMnAl-CoTiMnAl, CoRuMnAl-CoFeMnAl, CoRuMnSi-CoFeMnSi, the constituents with 4$d$ element as X$^{\prime}$ are all near half-metals while the other ones are all half-metals. Apart from the last two pairs, the components in the other three pairs have identical ground state structures, T$_{\text{I}}$. CoRuMnAl and CoRuMnSi crystallise in T$_{\text{II}}$ while CoFeMnAl and CoFeMnSi have T$_{\text{I}}$ as their ground states.

There are another eight pairs of isoelectronic compounds whose electronic behaviours are different. In each of these pairs compounds with 4$d$ elements are metals while the ones with all magnetic components beingf 3$d$ transition metals are half-metals. These pairs are CoNiMnAl-CoPdMnAl, CoMnFeAl-CoTcFeAl,CoCrFeSi-CoMoFeSi, CoMnFeSi-CoTcFeSi, CoCrMnSi-CoMoMnSi,CoVFeSi-CoNbFeSi,CoCrMnAl-CoMoMnAl and CoMn$_{2}$Al-CoTcMnAl. The constituents of the first five pairs have different ground state structures; the compounds with all 3$d$ magnetic components have structure type T$_{\text{I}}$ while the ones with one 4$d$ constituent are in structure type T$_{\text{II}}$. Out of the remaining three pairs, the constituents of the first two pairs crystallise in T$_{\text{I}}$ while the last one crystallise in T$_{\text{II}}$.

These observations clearly indicate that N$_{\text{V}}$ can be considered a predictor for half-metallic behaviour of quaternary Heuslers as long as structure types are same as that of existing isoelectronic half-metals. This was also observed in cases of ternary Heusler compounds \cite{srikrishna1}. The reason can be traced back to the similarities in the electronic structures and band-fillings for isoelectronic compounds across various different series with transition metal elements from different rows of the periodic table. Although the systematic analysis of evolution of properties with changes in N$_{\text{V}}$ through an analysis of evolutions in the partial densities of states providing a clear picture of element-specific hybridisations is not available in literature for quaternary compounds with all magnetic elements from 3$d$ row of periodic table, one can intuitively conclude that since the hybridisations in structure type T$_{\text{I}}$ compounds mostly come from the 3$d$ elements, the similarities in electronic structures for isoelectronic compounds across series and thus in the electronic properties like half-metallic behaviour are only natural consequences.

Another observation regarding the parents and daughter compounds in CoX$^{\prime}$Y$^{\prime}$Z considered here and in Reference \onlinecite{NSRP} is quite insightful. None of the quaternary daughters predicted to be half-metals among the compounds in these four series have both ternary parents as half-metals. The electronic properties of the ternary parents are computed in Reference \onlinecite{srikrishna1}. We find that out of the five half-metals CoYMnAl, CoNbFeAl, CoTcMnSi, CoRhMnSi and CoZrFeSi found in the quaternary series considered, Co$_{2}$YAl is a non-magnetic semiconductor, Mn$_{2}$YAl a metal; Co$_{2}$NbAl is a half-metal, Fe$_{2}$NbAl a non-magnetic semiconductor; both Co$_{2}$TcSi and Mn$_{2}$TcSi are normal metals; Co$_{2}$RhSi a metal while Mn$_{2}$RhSi a half-metal, Co$_{2}$ZrSi a half-metal and while Fe$_{2}$ZrSi is a non-magnetic semiconductor. This indicates that the half-metallicity in CoNbFeAl and CoZrFeSi might be coming solely from the half-metallicity of one of the ternary half-metal parents. This is because the other parent is a non-magnetic semiconductor and thus replacing one Fe with a Co in each one of them must have filled one of the spin bands rendering it metallic while the other spin band remains semiconducting. The origin of half-metallicity in the other three compounds would be indeed intriguing as there are three different combinations in their ternary parents: metal-metal, metal-non-magnetic semiconductor and metal-half-metal. This aspect is worth investigating in future.

%%%%%%%%%%%%%%%%%%%%%%%%%% End of Results and Discussions %%%%%%%%%%%%%%%%%
%================================================================================

%================================ Conclusion ===========================================
\section{Conclusions}

Using first-principles electronic structure calculations, we have systematically explored the structural, electronic and magnetic properties of  18 quaternary Heusler compounds, CoX$^\prime$Y$^\prime$Al where X$^\prime$  represents 9 elements with $4d$ electrons in their valance shells. We found two half-metallic ferromagnet CoYMnAl and CoRuMnAl with 100\% spin polarisation and three ''nearly'' half-metallic compounds with a gap like feature in minority spin channel and spin  polarisation greater than 90\%. The origin of half-metallicity or the lack of it in most of the compounds were explained using the changes in the electronic structures with number of valence electrons N$_{\text{V}}$. The trends in the magnetic moments, the inter-atomic exchange interactions and subsequently the magnetic transition temperatures T$_{\text{c}}$ with the changes in N$_{\text{V}}$ were found to be dependent upon the crystal structure of the compounds. The magnetic moments and magnetic exchange interactions could be correlated to the electronic structures. However, this work offered a broader perspective to the physics of half-metallic quaternary Heuslers. We systematically compared the results on CoX$^{\prime}$Y$^{\prime}$Al series with the existing results on another series with same Y$^{\prime}$ and X$^{\prime}$, CoX$^{\prime}$Y$^{\prime}$Si \cite{NSRP}. We found that as long as N$_{\text{V}}$ and structure types are the same, the electronic and magnetic properties of the compounds from the two series are nearly identical. The minor changes somewhat affecting the possibilities of half-metallic behaviour were due to the changes in the positions of the 3$d$ states with respect to the Fermi levels, an artefact of the deeper lying Si states pulling the $d$ bands towards lower energies a compared to the Al states. We further compared the impacts of the X$^{\prime}$ component on half-metallic behaviour by comparing the results of isoelectronic quaternary pairs where X$^{\prime}$ is a 3$d$ element in one case, and 4$d$ in the other. We once again observed that unless the structure types are identical, it is not possible to obtain half-metallicity in isoelectronic compounds from both series. Thus, rather than the chemical element X$^{\prime}$, the N$_{\text{V}}$ and the structure types are more important in determining half-metallic behaviour in quaternary Heuslers. We attributed this to the general picture of hybridisations between the transition metal atoms which were found to be transferrable across various quaternary series. Finally, we investigated the transferability of the half-metallic properties of parent ternary Heuslers making up a daughter quaternary compound. We found that neither of the quaternary Heusler compounds considered here and in Reference \onlinecite{NSRP} showing half-metallic behaviour have half-metallicity in both  parent ternary compounds. This poses an interesting open question regarding the emergence of half-metallic behaviour through substitution at select sites which will be addressed in the future. This work, therefore, provides a number of useful insights into evolving a broader picture of half-metallicity which can help predict and design future half-metals for spintronics applications. The importance of this work is that instead of focusing on prediction of new half-metals from a new series of compounds, which is what available mostly in literatures, it makes an attempt to address the microscopic physics governing a large group of materials in a systematic and coherent way. 
  
%%%%%%%%%%%%%%%%%%%%%%%%%% End of Conclusions %%%%%%%%%%%%%%%%%
%=======================================================================================
\section*{ACKNOWLEDGMENT}

We thank Dr. Ashis Kundu for his fruitful inputs and motivations through out this work. We acknowledge IIT Guwahati and DST India for providing the PARAM superconducting facility and the NEWTON computer cluster in the Department of Physics, IIT Guwhati.

\bibliographystyle{aip} % or  "apsrev4-1" "apsrmp4-1" "plain", "unsrt", "alpha", "abbrv", etc.
\bibliography{ref}

\end{document}